\documentclass[10pt,a4paper]{article}
\usepackage[english]{babel}
\usepackage[latin1]{inputenc}
\usepackage{amsfonts,amsbsy,bm,euscript,mathrsfs}
\usepackage{amssymb,stmaryrd,faktor}
\usepackage[tbtags]{amsmath}
\usepackage{bbm}
\usepackage{graphicx}
\usepackage[title,titletoc]{appendix}
\usepackage[bookmarks=true,colorlinks=true,linkcolor=blue,citecolor=blue,urlcolor=blue,bookmarksnumbered]{hyperref}

\numberwithin{equation}{section}

\makeatletter
\renewcommand\section{\@startsection {section}{1}{\z@}
{-3.5ex \@plus -1ex \@minus -.2ex}
{2.3ex \@plus.2ex}
{\normalfont\Large\bfseries}}
\renewcommand\subsection{\@startsection{subsection}{2}{\z@}
{-3.25ex\@plus -1ex \@minus -.2ex}
{1.5ex \@plus.2ex}
{\normalfont\large\bfseries}}
\makeatother

\def\smallL{\small{L}}
\def\smallR{\small{R}}
\newcommand{\alg}[1]{\mathfrak{#1}}

\newcommand{\gen}[1]{\mathfrak{#1}}

\begin{document}

\thispagestyle{empty}
\begin{flushright}\footnotesize\ttfamily
DMUS-MP-16/12
\end{flushright}
\vspace{2em}

\begin{center}

{\Large\bf \vspace{0.2cm}
{\color{black} $AdS_3/CFT_2$ and $q$-Poincar\'e superalgebras}} 
\vspace{1.5cm}

\textrm{\large Joakim Str\"omwall,\footnote{\texttt{j.stromvall@surrey.ac.uk}}
Alessandro Torrielli\footnote{\texttt{a.torrielli@surrey.ac.uk}}}

\vspace{2em}

\vspace{1em}
\begingroup\itshape
Department of Mathematics, University of Surrey,
\\ Guildford, GU2 7XH, UK
\par\endgroup

\end{center}

\vspace{2em}

\begin{abstract}\noindent 

\end{abstract}

We discover that a certain deformation of the $1+1$ dimensional Poincar\'e superalgebra is exactly realised in the massless sector of the $AdS_3/CFT_2$ integrable scattering problem. Deformed Poincar\'e superalgebras were previously noticed to appear in the $AdS_5/CFT_4$ correspondence - which displays only massive excitations -, but they were there only a partial symmetry. We obtain a representation of the boost operator and its coproduct, and show that the comultiplication exactly satisfies the homomorphism property. We present a classical limit, and finally speculate on an analogy with the physics of {\it phonons}.      

\newpage

\overfullrule=0pt
\parskip=2pt
\parindent=12pt
\headheight=0.0in \headsep=0.0in \topmargin=0.0in \oddsidemargin=0in

\vspace{-3cm}
\thispagestyle{empty}
\vspace{-1cm}

\tableofcontents

\setcounter{footnote}{0}

\section{Introduction}\label{secint}

\subsection{Introduction}

\paragraph{Unconventional quantum groups in the $AdS/CFT$ correspondence.}
The integrable structure underlying the $AdS/CFT$ correspondence \cite{Beisert:2010jr,Arutyunov:2009ga} keeps producing new exciting results which expand our understanding of exactly-solvable scattering problems and reveal new algebraic structures in quantum groups based on Lie superalgebras.    

The Hopf superalgebra which governs the $AdS_5/CFT_4$ integrable scattering problem is a rather exotic infinite-dimensional Yangian-type structure \cite{Drinfeld:1985rx,Molev,BYa}. The very nature of the Yangian implies that the algebra is layered (or {\it filtered}) into {\it levels}. The level-0 part is given by Beisert's centrally-extended $\alg{psu}(2|2)$ Lie superalgebra \cite{Beisert:2005tm,Arutyunov:2006ak}. At level 1, the Yangian charges are all paired-up with the corresponding level-0 ones, except for one extra generator \cite{Matsumoto:2007rh,Beisert:2007ty} which has no counterpart at level 0. This {\it secret} or {\it bonus} generator (hypercharge) $\mathfrak{B}$ is related to the fermion-number operator acting on the particles involved in the scattering process. The explicit form of Beisert's $S$-matrix \cite{Beisert:2005tm} allows for the scattering $|\mbox{boson} \rangle \otimes |\mbox{boson} \rangle \mapsto |\mbox{fermion} \rangle \otimes |\mbox{fermion} \rangle$ and vice versa, and therefore breaks the two-particle fermion number $(-)^F \otimes \mathbbm{1} + \mathbbm{1} \otimes (-)^F$ (with $F$ being the single-particle fermion-number operator). This symmetry is restored at Yangian-level 1 by means of a non-trivial {\it tail} added to the mere local coproduct-rule written above, according to a phenomenon typical of the higher quantum-group charges. Roughly speaking, what is conserved is
\begin{equation}
\label{secret}
\Delta(\mathfrak{B}) = {\mathfrak{b}}(p) \, (-)^F \otimes \mathbbm{1} + \mathbbm{1} \otimes {\mathfrak{b}}(p) \, (-)^F + \mathfrak{Q}^\alpha_a \, e^{i\frac{p}{4}} \otimes \alg{S}^a_\alpha \, e^{i\frac{p}{4}} + \alg{S}^a_\alpha \, e^{-i\frac{p}{4}} \otimes \mathfrak{Q}^\alpha_a \, e^{-i\frac{p}{4}},
\end{equation}
where the {\it tail} involves a complete contraction of all the supersymmetric indices, and $\mathfrak{b}(p)$ is a suitable function of the momentum.

The conservation of the level-zero hypercharge would immediately extend the algebra to $\alg{gl}(2|2)$, but this does not produce linear relations in the presence of the central extension. Nevertheless, recently a great progress \cite{Beisert:2014hya} has occurred, based on the $R\mathcal T\mathcal T$ formulation of the Yangian. In \cite{Beisert:2014hya}, in fact, it has been shown how the $S$-matrix generates, via the $R\mathcal T\mathcal T$ procedure, an abstract algebra, which reproduces all the relations of the $AdS_5$ integrable problem, including the {\it braiding element} as a generator at Yangian-level $-1$. These advances have revealed how the secret symmetry is embedded in the algebra, even though it is still a challenge to obtain the associated Drinfeld's second realisation \cite{Dsec}. Furthermore, the crossing-symmetry properties of the {\it secret} generator turn out to be particularly delicate. 

This phenomenon has by now been observed in a wide array of different sectors of the $AdS_5/CFT_4$ correspondence, including boundary problems \cite{Regelskis:2011fa}, scattering amplitudes \cite{Beisert:2011pn}, pure-spinor formulation \cite{Berkovits:2011kn}, quantum-affine deformations \cite{deLeeuw:2011fr} and Wilson loops \cite{Munkler}. This is a signal that such an occurrence is not an isolated property of the spectral problem \cite{deLeeuw:2012jf}. 

\paragraph{Other dimensions.} Similar exotic quantum-group structures and, in particular, the presence of the secret symmetry, have been found in $AdS/CFT$ correspondences formulated in different dimensionalities.

When it comes to the integrable scattering problem, the $AdS_4$ background has features which are very similar to the five-dimensional case, hence it does not seem to be particularly revealing on this specific issue.

The $AdS_3/CFT_2$ correspondence has been shown to lead to integrability \cite{Babichenko:2009dk} (for a review, see \cite{rev3}). The backgrounds have 16 supersymmetries and come in two main types: $AdS_3\times S^3\times T^4$ and $AdS_3\times S^3\times S^3\times S^1$. The latter has a continuum parameter $\alpha$ measuring the relative radii of the two three-spheres, reflected in the superisometry being embedded into the Lie superalgebra $\gen{D}(2,1;\alpha)\times \gen{D}(2,1;\alpha)$ as a superconformal algebra. By In\"on\"u-Wigner contraction $\alpha \to 0$ we obtain $\alg{psu}(1,1|2)\times\alg{psu}(1,1|2)$, which corresponds to the $T^4$-background. 

After classical integrability was shown \cite{Babichenko:2009dk} (see also \cite{Sundin:2012gc}), the finite-gap equations were derived and an all-loop Bethe ansatz was conjectured \cite{OhlssonSax:2011ms}. Initial explorations of the {\it massless modes} (which appear for $AdS/CFT$ in $3$ and $2$ dimensions\footnote{These modes are similar to a type of representations which were then later on studied in $AdS_5$ in \cite{Matsumoto:2014cka}. There, such representations are non-physical, and they have been dubbed {\it middle multiplets}.}) were reported in \cite{Sax:2012jv,Lloyd:2013wza}. These excitations were later fully incorporated in the world-sheet analysis of \cite{Borsato:2014exa,Borsato:2014hja} (see also \cite{Abbott:2014rca}). The massive scattering theory was first constructed in \cite{Borsato:2012ud,Borsato:2012ss} starting from a  centrally-extended algebra of the Beisert-type based on $\alg{psu}(1|1)$ factors. In \cite{Borsato:2013qpa}, the exact $S$-matrix for the $T^4$ case was obtained, with dressing phases being proposed in \cite{Borsato:2013hoa}. This $S$-matrix matched perturbative string predictions \cite{Rughoonauth:2012qd,Sundin:2012gc,Abbott:2012dd,Beccaria:2012kb,Beccaria:2012pm,Sundin:2013ypa,Bianchi:2013nra,Sundin:2016gqe}.   
On the contrary, certain mismatches with perturbation theory are present in the massless sector, concerning, in particular, the functional form of the dispersion-relation (see, for instance, the very recent \cite{Sundin:2016gqe,Borsato:2016xns}). Because the results of this paper crucially rely on the exact form of the massless dispersion-relation, our analysis will be particularly sensitive to these variations. Therefore, we hope that our alternative algebraic approach might actually be helpful to attack the problem of such mismatches from a new angle, and perhaps indicate a way of resolving them. 

Furthermore, recent ground-breaking work \cite{Sax:2014mea,Borsato:2016kbm} has demonstrated how the group-theoretical properties anticipated in \cite{Sax:2012jv} for the massless sector, are exactly reproduced by analysing a suitable field-theory dual, and the nature and symmetries of the associated matter content. For further work in the area, see \cite{Abbott:2013ixa,Sundin:2013uca,Borsato:2015mma,Prin,Abbott:2015mla,Per}, and for extensions to mixed R-R and NS-NS flux see \cite{Cagnazzo:2012se}. 

The secret symmetry has been found in \cite{Pittelli:2014ria}. It has subsequently been embedded in the more general framework of \cite{Regelskis:2015xxa}, where quantum deformations have also been studied. 

Going one dimension lower in the list of \cite{sym} one encounters the $AdS_2 \times S^2 \times T^6$ background with 8 supersymmetries. The dual field theory might either reduce to a superconformal quantum mechanics, or to a chiral CFT \cite{dual}. One has a
Metsaev-Tseytlin type action \cite{Metsaev:1998it,sc,Cagnazzo:2011at,Murugan:2012mf,amsw} for the coset part of the geometry, which is based on the supergroup
$$
\frac{PSU(1,1|2)}{ SO(1,1) \times SO(2)}~.
$$ 
Although one can truncate the classical action to its
coset component, there exists no choice of $\kappa$-symmetry gauge-fixing which may decouple the remaining fermions
\cite{Sorokin:2011rr}. The integrability of the
full background has been so far shown only to the quadratic order in the fermions
\cite{Sorokin:2011rr,Cagnazzo:2011at}.

In \cite{Hoare:2014kma} an exact $S$-matrix for the scattering of putative {\it magnon} excitations above the BMN vacuum
\cite{Berenstein:2002jq} has been proposed, relying on the same idea of centrally-extended residual symmetry algebras preserving the vacuum state. The light-cone gauge fixed Lagrangian
\cite{Murugan:2012mf,amsw} describes $2+2$ (bosons+fermions)
massive plus $6+6$ {\it massless} excitations. By deforming the coproduct in the familiar fashion \cite{tor} a massive-sector $S$-matrix was obtained, which satisfies crossing symmetry \cite{Janik:2006dc} and
unitarity, provided the dressing phase satisfies an appropriate set of relations. Preliminary consistency with perturbation theory was found \cite{Murugan:2012mf,amsw}. One of the main novelties of the two-dimensional case is the absence of a shortening condition which could allow to fix the dispersion-relation. The scattering in fact involves {\it long} representations \cite{Beisert:2006qh,Arutyunov:2009pw}. In \cite{Hoare:2014kma} the massless sector was also analysed and it was discovered that it is controlled by a canonical Yangian. 

The Yangian quantum group for the massive sector, and the associated secret symmetry, has been very recently found in \cite{Hoare:2015kla}.

\subsection{The present paper}

One of the striking features of the $AdS_3$ and $AdS_2$ integrability is without a doubt the presence of massless modes. Although, as we have mentioned above, they have by now been tamed into the Bethe ansatz, and a remarkable matching has been observed between their predicted group-theoretical properties and their {\it ``gauge-theory"} (respectively, their string-worldsheet) manifestation, it is certainly desirable to explore more fully the nature of their scattering and the consequences they have for the theory. This is particularly so when one tries to reconcile the matter with the expectation from standard integrable massless scattering \cite{Zamo} (see however \cite{Borsato:2016kbm} and the upcoming \cite{upcom}).

This paper attempts to analyse the issue of massless modes from a slightly different perspective, by probing their Hopf-algebra structure under a different angle: the one of deformed Poincar\'e superalgebras. In order to do that, we would first like to recall a series of observations which were put forward in the context of $AdS_5$, and which were then partly left unexplored. One  reason was that, in the $AdS_5$ case, such a deformed symmetry was only partially realised by the scattering problem. We intend to show that most of the properties pointed out in those early works are instead {\it exactly} realised in the {\it massless} sector of the $AdS_3$ superstring.

\paragraph{Quantum-deformed Poincar\'e supersymmetry}

In \cite{Gomez:2007zr}, the question was investigated whether any remnant of Poincar\'e symmetry could be found in the $AdS_5$ scattering problem. This was because relativistic invariance is traditionally rather powerful in constraining the $S$-matrix. The characteristic (massive) dispersion-relation of $AdS$ magnons was then interpreted as the Casimir of a certain $q$-deformation of the Poincar\'e algebra, and a suitable {\it boost} operator $\alg{J}$ was identified as acting as a shift on the elliptic rapidity-variable:
\begin{equation}
\alg{J}: z \to z + c.
\end{equation}
Subsequently, \cite{Charles} extended this picture to the centrally-extended $\alg{psu}(2|2)$ superalgebra, by matching it with a certain $q$-deformed super-Poincar\'e symmetry. This allowed for the reinterpretation of the specific supercharge-representation as a {\it boosted-frame} in terms of this deformed kinematical symmetry. However, an obstacle was pointed out insofar as some of the coproducts ceased to be symmetries of the $S$-matrix. A further mildly unnatural feature is also traditionally associated with the coproduct of the {\it energy} generator, and with its relationship to the one of the momentum, as we will later discuss. 

Afterwards, $q$-deformations of the $AdS_3$ Hopf-algebra structure have appeared in \cite{Hoare:2011fj} in the context of the Pohlmeyer-reduction approach\footnote{In the $AdS_5$ case, a parallel development connected to quantum-group deformations originated in \cite{BeisertK} (see also \cite{BenStijn,Sfetsos:2013wia}, which initiated much recent work, and \cite{Riccardo} for a review). In \cite{Pachol:2015mfa}, the $\eta$-deformed $AdS_5$ model \cite{Riccardo} has been connected {\it via} In\"on\"u-Wigner contraction to a deformation of the flat-space superstring, exhibiting $q$-deformed Poincar\'e symmetry. By the results of \cite{Arutyunov:2014cra} (see also \cite{Arutynov:2014ota}), the latter is then put in relationship with the $AdS_5$ {\it mirror model} \cite{Arutyunov:2007tc}, giving an alternative interpretation to the off-shell central extension of the $AdS_5 \times S^5$ string light-cone symmetry in terms of the $q$-deformed super-Poincar\'e algebra. Complications with the fermionic degrees of freedom \cite{Arutyunov:2014jfa,ArutyunovBorsatoFrolov} currently make this an open avenue for investigation (we thank S. van Tongeren for communication about this connection). In these and other contemporary lines of investigation, however, the quantum deformation is often super-imposed to the traditional one. This is not what we consider here, where the $q$-Poincar\'e deformation rather amounts to a co-existing alternative picture. Let us also notice that boost operators on long-range spin-chains of the type appearing in the $AdS/CFT$ correspondence have been studied in \cite{Bargheer:2008jt}, and they recently featured in the work of \cite{Beisert:2016qei}.}.

What we show in this paper is that, because of the peculiar nature of the massless dispersion-relation, such a deformed algebra actually becomes an {\it exact symmetry} of the massless $AdS_3$ $S$-matrix, when particles are taken to live on the same branch of the dispersion-relation. This gives in such a circumstance an alternative algebraic description of the massless scattering problem - companion to the one which is already in place.
Our hope is that this new picture might be more suggestive of the true nature of the massless modes as it is, in some sense, closer to a relativistic interpretation. This might eventually make it closer to the treatment of \cite{Zamo}. In a sense (to be made precise in the next section), the new coproduct is more natural than the pre-existing one, allowing for a single rule to be consistently applied across all the central charges to obtain their corresponding two-particle action.

\paragraph{Plan of the paper}

Section 2 will define the $q$-deformed algebra and coalgebra, the appropriate representation, and will recall the $R$-matrix. We shall discuss the deformed Casimir and its relationship to the massless dispersion-relation. We will then equip the boost generator with a suitable coproduct, and give a complete proof that it is an algebra homomorphism. We will also construct a classical limit and its associated classical algebra {\it via} a scaling procedure.

Section 3 will provide some intriguing physical analogies with the dynamics of phonons and {\it umklapp scattering}. Section 4 shall discuss how the boost generator is represented in a way which uniformises the dispersion-relation. 

Section 5 will finally provide some comments and future directions of investigation. 

\section{Deformations of Poincar\'e as invariances of the massless $S$-matrix}

Let us begin by writing the action of the symmetry generators on the elementary {\it massless} excitations, by focusing on the centrally-extended algebra associated to $\alg{su}(1|1)_L \oplus \alg{su}(1|1)_R $. Our algebraic analysis will be independent on whether this describes scattering in the $S^3 \times S^1$ rather than the $T^4$ geometry (where one shall take two copies of the same algebra). The non-vanishing (anti-)commutation relations read
\begin{equation}
\label{alge}
\{\alg{Q}_{\smallL}, \alg{S}_{\smallL}\} \ = \ \alg{H}_{\smallL}, \quad \{\alg{Q}_{\smallR}, \alg{S}_{\smallR}\} \ = \ \alg{H}_{\smallR}, \quad \{\alg{Q}_{\smallL}, \alg{Q}_{\smallR}\} \ = \alg{P}~, \qquad  \{\alg{S}_{\smallL}, \alg{S}_{\smallR}\} \ = \ \alg{K}~. 
\end{equation}
Restricting to {\it left-left} scattering \cite{Borsato:2014exa,Borsato:2014hja}, the representation we consider is described by a doublet $\{|\phi\rangle, |\psi\rangle\}$ with symmetry action given by
\begin{equation}\label{leftrep}
\begin{gathered}
\alg{S}_{\smallR} \ := \ - \sqrt{h \sin \frac{p}{2}}\begin{pmatrix}0&0\\1&0\end{pmatrix}~,\qquad
\alg{Q}_{\smallR} \ := \ - \sqrt{h \sin \frac{p}{2}}\begin{pmatrix}0&1\\0&0\end{pmatrix}, \\
\alg{Q}_{\smallL} \ := \ \sqrt{h \sin \frac{p}{2}}\begin{pmatrix}0&0\\1&0\end{pmatrix}~,\qquad
\alg{S}_{\smallL} \ := \ \sqrt{h \sin \frac{p}{2}}\begin{pmatrix}0&1\\0&0\end{pmatrix}. 
\end{gathered}
\end{equation}
We have chosen for the remainder of this paper a suitable branch of the dispersion-relation, which we will constrain to be {\it the same} for both scattering particles\footnote{Our choice will be the positive branch throughout:
\begin{equation}
x^\pm = e^{\pm i \frac{p}{2}}.
\end{equation}
As it is argued in \cite{Borsato:2014exa,Borsato:2014hja}, the scattering is still meaningful in this case, as the group-velocity is less than the speed of light (non-relativistic setting). If we take real momentum, this amounts to considering $p \in [0, \pi]$. We would like to thank A. Sfondrini for pointing out that our analysis might extend to include negative-branch particles, suitably selecting the fundamental region of momenta \cite{upcom}. This is because of the $2 \pi$-periodicity effectively generated in the algebra representation, cf. $E = h \big|\sin \frac{p}{2}\big|$. It will be interesting to construct the generalisation to our choice of coalgebra structure. We plan to pursue this idea in future work.}, such that the eigenvalues of the central charges are given by  
\begin{equation}\label{eq:defofex}
H_{\smallL} = H_{\smallR} = - P = - K = h \, \sin \frac{p}{2}.
\end{equation}
The quantity $h$ is the coupling constant of the theory. 

It is known \cite{Borsato:2013qpa} that there exists a coproduct (and in fact, an entire Yangian tower of symmetries) one can put on the algebra (\ref{alge}), which reproduces, when combined with the representation (\ref{leftrep}), the scattering matrices for the massive (and, by reduction \cite{upcom}, massless) $AdS_3$ sector. One can write this {\it old} coproduct as
\begin{equation}
  \label{coprodold}
  \begin{aligned}
    \Delta(\alg{H}_{\smallL})\ :=\ \alg{H}_{\smallL} \otimes \mathbbm{1} + \mathbbm{1} \otimes \alg{H}_{\smallL}, \qquad&
    \Delta(\alg{H}_{\smallR})\ :=\ \alg{H}_{\smallR} \otimes \mathbbm{1} + \mathbbm{1} \otimes \alg{H}_{\smallR}~, \\
     \Delta(\alg{Q}_{\smallL})\ :=\ \alg{Q}_{\smallL} \otimes {e^{-i \frac{p}{4}}} + {e^{i \frac{p}{4}}} \otimes \alg{Q}_{\smallL}, \qquad&
     \Delta(\alg{S}_{\smallL})\ :=\ \alg{S}_{\smallL} \otimes {e^{i \frac{p}{4}}} + {e^{-i \frac{p}{4}}} \otimes \alg{S}_{\smallL}~,\\
     \Delta(\alg{Q}_{\smallR})\ :=\ \alg{Q}_{\smallR} \otimes {e^{-i \frac{p}{4}}} + {e^{i \frac{p}{4}}} \otimes \alg{Q}_{\smallR}, \qquad&
     \Delta(\alg{S}_{\smallR})\ :=\ \alg{S}_{\smallR} \otimes {e^{i \frac{p}{4}}} + {e^{-i \frac{p}{4}}} \otimes \alg{S}_{\smallR}~,\\
      \Delta(\alg{P})\ :=\ \alg{P} \otimes {e^{-i \frac{p}{2}}} + {e^{i \frac{p}{2}}} \otimes \alg{P}, \qquad&
      \Delta(\alg{K})\ :=\ \alg{K} \otimes {e^{i \frac{p}{2}}} + {e^{-i \frac{p}{2}}} \otimes \alg{K},
  \end{aligned}
\end{equation}
where $p$ is the momentum operator, which is central in $\alg{su}(1|1)_L \oplus \alg{su}(1|1)_R $, and has trivial coproduct $\Delta(p) = p \otimes \mathbbm{1} + \mathbbm{1} \otimes p$.
Letting $\Phi$ be the overall scalar factor, it can be checked that the $R$-matrix defined by
\begin{equation}\label{eq:RLL}
  \begin{aligned}
    R |\phi\rangle \otimes |\phi\rangle\ &:=\ {\Phi} \, |\phi\rangle \otimes |\phi\rangle, \\
    R |\phi\rangle \otimes |\psi\rangle\ &:=\ - {\Phi} \, \csc \frac{p_1 + p_2}{4} \, \sin \frac{p_1 - p_2}{4} |\phi\rangle \otimes |\psi\rangle + {\Phi} \, \csc \frac{p_1 + p_2}{4} \, \sqrt{\sin \frac{p_1}{2} \sin \frac{p_2}{2}} |\psi\rangle \otimes |\phi\rangle, \\
    R |\psi\rangle \otimes |\phi\rangle\ &:=\ {\Phi} \, \csc \frac{p_1 + p_2}{4} \, \sin \frac{p_1 - p_2}{4}  |\psi\rangle \otimes |\phi\rangle + {\Phi} \, \csc \frac{p_1 + p_2}{4} \, \sqrt{\sin \frac{p_1}{2} \sin \frac{p_2}{2}} |\phi\rangle \otimes |\psi\rangle, \\
    R |\psi\rangle \otimes |\psi\rangle\ &:=\ - {\Phi} \, |\psi\rangle \otimes |\psi\rangle,
\end{aligned}
\end{equation}
indeed satisfies the invariance equation
\begin{equation}\label{definiLL}
  \Delta^{\text{op}} (\mathfrak{a})\,  R\ =\ R\, \Delta (\mathfrak{a}) \, \, \, \, \forall \, \, \mathfrak{a} \in \alg{sl}(1|1)_{\smallL}\oplus \alg{sl}(1|1)_{\smallR}.
\end{equation}
In this formula, $\Delta^{op} = \Pi (\Delta)$, with $\Pi$ being the graded permutation on the tensor-product algebra. This $R$-matrix also satisfies the Yang-Baxter equation by direct check:
\begin{equation}
R_{12} \, R_{13} \,R_{23} \, = R_{23} \,R_{13} \,R_{12}.
\end{equation}
It also satisfies the unitarity condition $\Pi(R)(p_2,p_1) \, R(p_1,p_2) = \mathbbm{1} \otimes \mathbbm{1}$ (if the scalar factor $\Phi$ satisfies an appropriate functional equation \cite{upcom})\footnote{We also notice an interesting fact. When stripped off of the scalar factor (meaning, when normalised in such a way that the purely bosonic entry equals $1$), the $R$-matrix satisfies $\Pi(R)(p_2,p_1) = R(p_1,p_2)$ and $R(p_1,p_2)^2 = \mathbbm{1} \otimes \mathbbm{1}$.}.

What we find in this paper is that there exists \underline{\it another} coproduct admitted by the same algebra, under which the {\it very same} $S$-matrix of the massless $AdS_3$ problem is also invariant. This means that the same invariance equation holds for the \underline{\it new} coproduct $\Delta_N$, which we endow the same algebra with:
\begin{equation}
  \label{coprod}
  \begin{aligned}
    \Delta_N(\alg{P})\ :=\ \alg{P} \otimes e^{i \frac{p}{2}} + e^{-i \frac{p}{2}} \otimes \alg{P}, \qquad&
    \Delta_N(\alg{K})\ :=\ \alg{K} \otimes e^{i \frac{p}{2}} + e^{-i \frac{p}{2}} \otimes \alg{K}~, \\
     \Delta_N(\alg{Q}_{\smallL})\ :=\ \alg{Q}_{\smallL} \otimes e^{i \frac{p}{4}} + e^{-i \frac{p}{4}} \otimes \alg{Q}_{\smallL}, \qquad&
     \Delta_N(\alg{S}_{\smallL})\ :=\ \alg{S}_{\smallL} \otimes e^{i \frac{p}{4}} + e^{-i \frac{p}{4}} \otimes \alg{S}_{\smallL}~,\\
     \Delta_N(\alg{Q}_{\smallR})\ :=\ \alg{Q}_{\smallR} \otimes {e^{i \frac{p}{4}}} + {e^{-i \frac{p}{4}}} \otimes \alg{Q}_{\smallR}, \qquad&
     \Delta_N(\alg{S}_{\smallR})\ :=\ \alg{S}_{\smallR} \otimes {e^{i \frac{p}{4}}} + {e^{-i \frac{p}{4}}} \otimes \alg{S}_{\smallR}~,\\
      \Delta_N(\alg{H}_{\smallR})\ :=\ \alg{H}_{\smallR} \otimes {e^{i \frac{p}{2}}} + {e^{-i \frac{p}{2}}} \otimes \alg{H}_{\smallR}, \qquad&
      \Delta_N(\alg{H}_{\smallL})\ :=\ \alg{H}_{\smallL} \otimes {e^{i \frac{p}{2}}} + {e^{-i \frac{p}{2}}} \otimes \alg{H}_{\smallL}~,\\ \Delta_N(p) = p \otimes \mathbbm{1} + \mathbbm{1} \otimes p
  \end{aligned}
\end{equation}
such that
\begin{equation}\label{definiLL2}
  \Delta_N^{\text{op}} (\mathfrak{a})\,  R\ =\ R\, \Delta_N (\mathfrak{a}) \, \, \, \, \forall \, \, \mathfrak{a} \in \alg{sl}(1|1)_{\smallL}\oplus \alg{sl}(1|1)_{\smallR}~.
\end{equation}
Following \cite{Gomez:2007zr,Charles}, we put this in correspondence with two copies of the 1+1 dimensional {\it $q$-deformed super-Poincar\'e algebra}, associated to $\alg{sl}(1|1)$, where
\begin{equation}
q \equiv e^{i h^{-2}}.
\end{equation}
We define an individual copy - which we will call $\alg{E}_q(1,1)$ - as follows:
\begin{eqnarray}
\label{algeq1}
&&\{\alg{q}, \alg{s}\} \ = \ \alg{h}, \quad [\alg{J},p] \ = \ i \alg{h}, \quad [\alg{J}, \alg{h}] \ = \ \frac{e^{i p} - e^{-i p}}{2 \mu}, \nonumber \\
&&[\alg{J}, \alg{q}] = \frac{i}{2 \sqrt{\mu}} \frac{e^{i \frac{p}{2}} + e^{- i \frac{p}{2}}}{2} \, \alg{q}, \qquad [\alg{J}, \alg{s}] = \frac{i}{2 \sqrt{\mu}} \frac{e^{i \frac{p}{2}} + e^{- i \frac{p}{2}}}{2} \, \alg{s},
\end{eqnarray}
treating $p$ as an additional independent generator, and having defined
\begin{equation}
\mu \equiv \frac{4}{h^2}
\end{equation}
in terms of the coupling constant $h$ of the model.
When we combine the two copies in $\alg{E}_q(1,1)_{\smallL} \oplus \alg{E}_q(1,1)_{\smallR}$, we identify
\begin{equation}
p_{\smallL} =p_{\smallR} \equiv p
\end{equation}
and write
\begin{eqnarray}
\label{algeq}
&&\{\alg{Q}_{\smallR}, \alg{S}_{\smallR}\} \ = \ \alg{H}_{\smallR}, \quad \{\alg{Q}_{\smallL}, \alg{S}_{\smallL}\} \ = \ \alg{H}_{\smallL}, \quad [\alg{J}_{\smallR}, p] \ = \ i \alg{H}_{\smallR}, \quad [\alg{J}_{\smallL}, p] \ = \ i \alg{H}_{\smallL},\nonumber \\ 
&&\qquad \qquad \qquad \qquad \qquad [\alg{J}_A, \alg{H}_B] \ = \frac{e^{i p} - e^{-i p}}{2 \mu},\nonumber\\
&&[\alg{J}_A, \alg{Q}_B] \ = \frac{i}{2 \sqrt{\mu}} \frac{e^{i \frac{p}{2}} + e^{- i \frac{p}{2}}}{2} \, \alg{Q}_B, \qquad [\alg{J}_A, \alg{S}_B] = \frac{i}{2 \sqrt{\mu}} \frac{e^{i \frac{p}{2}} + e^{- i \frac{p}{2}}}{2} \, \alg{S}_B,
\end{eqnarray}
where {\footnotesize $(A,B) = (L,L), (L,R), (R,L), (R,R)$}. We also centrally-extend the resulting algebra according to (\ref{alge}):
\begin{eqnarray}
\label{algeqc}
&&\{\alg{Q}_{\smallL}, \alg{Q}_{\smallR}\} \ = \ \alg{P}~, \qquad  \{\alg{S}_{\smallL}, \alg{S}_{\smallR}\} \ = \ \alg{K},\nonumber \\ 
&&[\alg{J}_{\smallL}, \alg{P}] \ = [\alg{J}_{\smallR}, \alg{P}] \ = \ [\alg{J}_{\smallL}, \alg{K}] \ = [\alg{J}_{\smallR}, \alg{K}]= \frac{e^{- i p} - e^{i p}}{2 \mu}.
\end{eqnarray}
We will discuss the meaning of the generator $\alg{J}$ in what follows.

Let us reproduce the argument of \cite{Gomez:2007zr}, now for massless modes. The combination 
\begin{eqnarray}
\alg{C}_2 \equiv \frac{\alg{H}^2}{4} + \frac{1}{\mu} \, (e^{i p} + e^{-i  p} - 2), \qquad \alg{H} = \alg{H}_{\smallL} + \alg{H}_{\small_R},
\end{eqnarray}
commutes with all the generators of the centrally-extended $\alg{E}_q(1,1)_{\smallL} \oplus \alg{E}_q(1,1)_{\smallR}$ superalgebra, in particular 
\begin{equation}
[\alg{J}_{\small_R}, \alg{C}_2] =[\alg{J}_{\small_L}, \alg{C}_2] =0
\end{equation}
simply from the algebra relations. In our representation, setting to zero the eigenvalue of this central element reproduces the massless dispersion-relation:
\begin{equation}
C_2 = H^2 - 4 h^2 \sin^2 \frac{p}{2} = 0, \qquad H = H_{\smallL} + H_{\smallR}.
\end{equation}

The undeformed limit is obtained by keeping the supercharges finite, and scaling
\begin{eqnarray}
\label{unde}
p \to \epsilon p_1, \qquad h \to \frac{c}{\epsilon}, \qquad \alg{J}_A \to \frac{i \alg{b}}{2 \epsilon}, \qquad H \to e_0, \qquad \epsilon \to 0,
\end{eqnarray}
implying $q \to 1$,
\begin{eqnarray}
[\alg{b},p_1] = e_0, \qquad [\alg{b},e_0] = c^2 p_1, \qquad [\alg{b},\alg{Q}_A] = \frac{c}{2} \alg{Q}_A, \qquad [\alg{b},\alg{S}_A] =\frac{c}{2} \alg{S}_A, \qquad [\alg{b},\alg{P}] =  [\alg{b},\alg{K}] = - c^2 p_1, \nonumber
\end{eqnarray}
and the relativistic dispersion
\begin{eqnarray}
e_0^2 - c^2 p_1^2 =0.
\end{eqnarray}

\medskip

Let us conclude this section with a few remarks.

First, one can check that the Jacobi identity is satisfied for the algebra-relations we have postulated. This will actually be specified more clearly in the next section.

Furthermore, we would like to argue that the new coproduct 
$\Delta_N$ has a remarkable property, which makes it, in a sense, more natural than the traditional one {\it for the massless sector}. Because of the peculiar nature of the massless dispersion-relation, it is only for massless particles, and for the new coproduct, that the comultiplication map for the {\it energy generator} $\mathfrak{H}$ actually {\it follows} from the one postulated for the momentum generator $p$. This will be demonstrated more clearly in the next section. In no other circumstances is this the case, since the square-root formula for the energy always prevents this occurrence, forcing one to postulate the coproduct of the energy separate from the momentum, in order to close the algebra. In the massive case, in fact, one normally has to prescribe
\begin{eqnarray}
\Delta\Big({\cal{H}}(p)\Big) \to H(p_1) + H(p_2), \qquad H_i = \sqrt{m_i^2 + 4 h^2 \sin^2 \frac{p_i}{2}}    
\end{eqnarray}
which, since $\Delta(p) \to p_1 + p_2$, necessarily implies
\begin{eqnarray}
\Delta\Big({\cal{H}}(p)\Big) \neq {\cal{H}}\Big(\Delta(p)\Big).
\end{eqnarray}
One needs the specific deformation $\Delta_N$ of the coalgebra structure, and the precise form of the $\sin \frac{p}{2}$ function, both conspiring to remove this slightly unnatural feature, and obtain
\begin{eqnarray}
\Delta_N\Big({\cal{H}}(p)\Big) = {\cal{H}}\Big(\Delta_N(p)\Big).
\end{eqnarray}
This is of course at the price of loosing the interpretation of $\Delta_N({\cal{H}})$ as the total energy $H_1+H_2$, as we will discuss in section \ref{phys}. Nevertheless, the charge $\Delta_N({\cal{H}})$ is conserved in the scattering process.

Furthermore, the {\it old} comultiplication is best suited to describe the Yangian tower of integrable charges, for which we have no analogue at all at the moment in the picture we are describing here.

\subsection{Coproduct for Boosts}

In this section we want to show that the coproduct given in \cite{Charles} for the boost operator $\alg{J}$ is an exact Lie-algebra homomorphism of our deformed algebra. This means that the coproducts should satisfy the same defining relations of the algebra, signifying that all the (anti-)commutation relations are realised also on two-particle states. In accordance with the discussion at the end of the previous section, we will show that this extends to {\it any} central charge which is a function of the momentum, whose coproduct is then systematically obtained by evaluating the function itself on $\Delta_N(p)$.

Let us focus on the $L$ sector\footnote{It turns out that the expression (\ref{deltaJ}) will work for the $R$ sector as well. This is due to the isomorphism between the $L$ and $R$ representations, which in turn is easily seen from the relations given in the previous subsection.}. The coproduct for the boost operator in this sector is given by a similar expression as in \cite{Charles}, adapted to the $AdS_3$ massless $L$ algebra:
\begin{equation}
\label{deltaJ}
\Delta_N (\alg{J}_L) = \alg{J}_L \otimes e^{i \frac{p}{2}} +  e^{-i \frac{p}{2}} \otimes \alg{J}_L + \frac{1}{2} \, \alg{Q}_L \, e^{- i \frac{p}{4}} \otimes \alg{S}_L \, e^{i \frac{p}{4}} +  \frac{1}{2} \, \alg{S}_L \, e^{- i \frac{p}{4}} \otimes \alg{Q}_L \, e^{i \frac{p}{4}}.
\end{equation}

{\it Proof.}

\medskip

Let us start by showing the need for the non-trivial phases in the first two terms of (\ref{deltaJ}) - namely, those terms where the boost-operator itself is present. These phases can be fixed by requiring the homomorphism property 
\begin{equation}
[\Delta_N (\alg{J}_L), \Delta_N (p)] = \Delta_N (\alg{H}_L).
\end{equation}
This is because
\begin{equation}
[\Delta_N (\alg{J}_L), \Delta_N (p)] = [\alg{J}_L, p] \otimes  e^{i \frac{p}{2}} +  e^{-i \frac{p}{2}} \otimes [\alg{J}_L, p], 
\end{equation}
which can be seen to coincide with $i$ times the coproduct $\Delta_N (\alg{H}_L)$ given in (\ref{coprod}).  
The {\it tail} of (\ref{deltaJ}) (namely, the part featuring the supercharges) does not contribute to this commutator, since $p$ commutes with the supercharges. Similarly, the tail does not contribute to 
\begin{equation}
[\Delta_N (\alg{J}_L), \Delta_N (\alg{H}_L)] = \Delta_N \Big( (2 \mu)^{-1} \, (e^{i p} - e^{-i p})\Big)
\end{equation}
either. However, here the boost generator also acts on the momentum-dependent phase factors, and one obtains four terms:
\begin{equation}
[\Delta_N (\alg{J}_L), \Delta_N (\alg{H}_L)] = [\alg{J}_L, \alg{H}_L] \otimes  e^{i p} +  [\alg{J}_L, e^{-i \frac{p}{2}}] \otimes  e^{i \frac{p}{2}} \alg{H}_L + e^{-i \frac{p}{2}} \alg{H}_L \otimes [\alg{J}_L, e^{i \frac{p}{2}}] + e^{-i p} \otimes [\alg{J}_L, \alg{H}_L]. 
\end{equation}
By using Taylor expansion, and the fundamental commutator $[\alg{J}_L,p] = i \alg{H}_L$, one can prove that
\begin{equation}
[\alg{J}_L,e^{i \alpha p}] = - \alpha \, \alg{H}_L \, e^{i \alpha p}
\end{equation}
for any constant $\alpha$, by which we obtain
\begin{equation}
\label{rhs}
[\Delta_N (\alg{J}_L), \Delta_N (\alg{H}_L)] = \frac{1}{2 \mu} (e^{i p} \otimes e^{i p} - e^{-i p} \otimes e^{-i p}), 
\end{equation}
where we have used that 
\begin{equation}
\label{ow}
\Delta_N (e^{i \alpha p}) = e^{i \alpha p} \otimes e^{i \alpha p}.
\end{equation}

\medskip

For the commutator with the supercharges the calculation is more involved, since the tail of (\ref{deltaJ}) contributes non-trivially. To proceed, we first work with a general set of Lie-algebra relations
\begin{equation}
[\alg{J}_L, \alg{Q}_L] = \phi(p) \, \alg{Q}_L, \qquad [\alg{J}_L, \alg{S}_L] = \tilde{\phi}(p) \, \alg{S}_L,
\end{equation}
with functions $\phi(p)$ and $\tilde{\phi}(p)$ for the moment left unspecified. A non-trivial Jacobi identity for the algebra given in the previous section reads\footnote{Notice that, for instance, Jacobi identities of the type $[\alg{J}, [\mbox{central}, \alg{Q}]]$ are in principle non-trivial, however they end up being identically realised because of the centrality of $[\alg{J}, \mbox{central}]$. The same holds for $\alg{S}$. The only other identities to be checked are then all the remaining ones of the form $[\alg{J}, \{\alg{Q}_A,\alg{S}_B\}]$ with $A,B$ all the other possible combinations of $L,R$. These are either identically zero, or they reduce to (\ref{Jac}).}
\begin{equation}
\label{Jac}
\frac{i \sin p}{\mu} = [\alg{J}_L, \alg{H}_L] = [\alg{J}_L, \{\alg{Q}_L,\alg{S}_L\}] = \{ [\alg{J}_L,\alg{Q}_L], \alg{S}_L\} + \{ [\alg{J}_L,\alg{S}_L], \alg{Q}_L\} = \Big(\phi(p) + \tilde{\phi}(p)\Big) \, \alg{H}_L,
\end{equation} 
from which one gets the constraint
\begin{equation}
\label{Jacob}
\phi(p) + \tilde{\phi}(p) = \frac{i \sin p}{H_L \, \mu} \, = \, \frac{i h}{2} \cos \frac{p}{2}. 
\end{equation}
We also work with a generic expression 
\begin{equation}
\label{deltaJgen}
\Delta_N (\alg{J}_L) = \alg{J}_L \otimes e^{i \frac{p}{2}} +  e^{-i \frac{p}{2}} \otimes \alg{J}_L + A(p_1,p_2) \, \alg{Q}_L \, \otimes \alg{S}_L \,  +  B(p_1,p_2) \, \alg{S}_L \, \otimes \alg{Q}_L,
\end{equation}
and let consistency fix the unknown functions $A(p_1,p_2)$ and $B(p_1,p_2)$.

Imposing now the homomorphism property amounts to the following two conditions:
\begin{equation}
\label{sys}
[\Delta_N (\alg{J}_L), \Delta_N (\alg{Q}_L)] = \Delta_N \Big(\phi(p)\Big) \, \Delta_N (\alg{Q}_L), \qquad  [\Delta_N (\alg{J}_L), \Delta_N (\alg{S}_L)] = \Delta_N \Big(\tilde{\phi}(p)\Big) \, \Delta_N (\alg{S}_L).
\end{equation}
By keeping track of the terms of the type with $\alg{Q}_L$ and $\alg{S}_L$ in the first and second factor of the tensor product, respectively, we can reduce the matrix-equations (\ref{sys}) to the following system of four scalar equations:
\begin{eqnarray}
&&\phi(p_1) \, e^{i \frac{3}{4} p_2} - \frac{1}{4} \, H_{L,2} \, e^{i \frac{1}{4} p_2 - i \frac{1}{2} p_1} + A(p_1,p_2) \, \alg{H}_{L,2} \, e^{-i \frac{1}{4} p_1} = \phi_{12}  \,  \, e^{i \frac{1}{4} p_2},\nonumber \\
&&\phi(p_2) \, e^{-i \frac{3}{4} p_1} + \frac{1}{4} \, H_{L,1} \, e^{i \frac{1}{2} p_2 - i \frac{1}{4} p_1} - B(p_1,p_2) \, \alg{H}_{L,1} \, e^{i \frac{1}{4} p_2} = \phi_{12}  \,  \, e^{-i \frac{1}{4} p_1},\nonumber \\
&&\tilde{\phi}(p_1) \, e^{i \frac{3}{4} p_2} - \frac{1}{4} \, H_{L,2} \, e^{i \frac{1}{4} p_2 - i \frac{1}{2} p_1} + B(p_1,p_2) \, \alg{H}_{L,2} \, e^{-i \frac{1}{4} p_1} = \tilde{\phi}_{12}  \,  \, e^{i \frac{1}{4} p_2},\nonumber \\
&&\tilde{\phi}(p_2) \, e^{-i \frac{3}{4} p_1} + \frac{1}{4} \, H_{L,1} \, e^{i \frac{1}{2} p_2 - i \frac{1}{4} p_1} - A(p_1,p_2) \, \alg{H}_{L,1} \, e^{i \frac{1}{4} p_2} = \tilde{\phi}_{12}  \,  \, e^{-i \frac{1}{4} p_1},
\end{eqnarray}  
where the symbols $\phi_{12}$ and $\tilde{\phi}_{12}$ represent the eigenvalues of the corresponding respective coproducts $\Delta_N \Big(\phi(p)\Big)$  and $\Delta_N \Big(\tilde{\phi}(p)\Big)$ (which are central in the tensor-product algebra, hence proportional to the $4 \times 4$ identity matrix).  

After a rather lenghty calculation and numerous simplifications owing to trigonometric identities, the following solution can be found:
\begin{eqnarray}
&&B(p_1,p_2) = \frac{\Big[\frac{3}{4}\, H_{L,2} - \phi(p_2)\Big]\, e^{-i \frac{1}{4} (p_1+p_2)}-\Big[\frac{1}{4}\, H_{L,1} - \phi(p_1)\Big]\, e^{i \frac{1}{4} (p_1+p_2)}}{H_{L,2} \, e^{-i \frac{1}{2} p_2} - H_{L,1} \, e^{i \frac{1}{2} p_1}},\nonumber \\ 
&&A(p_1,p_2) = \frac{\Big[\frac{1}{4}\, H_{L,2} + \phi(p_2)\Big]\, e^{-i \frac{1}{4} (p_1+p_2)}-\Big[\frac{3}{4}\, H_{L,1} + \phi(p_1)\Big]\, e^{i \frac{1}{4} (p_1+p_2)}}{H_{L,2} \, e^{-i \frac{1}{2} p_2} - H_{L,1} \, e^{i \frac{1}{2} p_1}},
\end{eqnarray}
and
\begin{eqnarray}
&&\phi_{12} = \frac{\frac{1}{2} \, H_{L,1} \, H_{L,2} + \phi(p_1) \, H_{L,1} \, e^{i \frac{(p_1+p_2)}{2}} - \phi(p_2) \, H_{L,2} \, e^{-i \frac{(p_1+p_2)}{2}}}{H_{L,1} \, e^{i \frac{1}{2} p_1} - H_{L,2} \, e^{-i \frac{1}{2} p_2}}\nonumber \\
&&\tilde{\phi}_{12} = \frac{\frac{1}{2} \, H_{L,1} \, H_{L,2} + \tilde{\phi}(p_1) \, H_{L,1} \, e^{i \frac{(p_1+p_2)}{2}} - \tilde{\phi}(p_2) \, H_{L,2} \, e^{-i \frac{(p_1+p_2)}{2}}}{H_{L,1} \, e^{i \frac{1}{2} p_1} - H_{L,2} \, e^{-i \frac{1}{2} p_2}}.
\end{eqnarray}

We can now impose $\phi$ and $\tilde{\phi}$ to be equal to each other, which, given the constraint (\ref{Jacob}) imposed by the Jacobi identity, sets them both equal to
\begin{equation}
\phi(p) = \tilde{\phi}(p) = \frac{i h}{4} \cos \frac{p}{2}.
\end{equation}
This matches with the commutation relations we have given in the previous subsection. With this choice, one observes a dramatic simplification of the above results, which reduces to
\begin{equation}
\label{expre}
A(p_1,p_2) = B(p_1,p_2) = \frac{1}{2} e^{i \frac{(p_2 - p_1)}{4}}, \qquad \phi_{12} = \tilde{\phi}_{12} = \frac{i h}{4} \cos \frac{p_1 + p_2}{2}.
\end{equation}
The above formulas not only produce the form of the coproduct displayed in (\ref{deltaJ}), but they also show that consistency is achieved. In fact, the expression (\ref{expre}) for the eigenvalue of the coproducts  $\Delta_N \Big(\phi(p)\Big)$  and $\Delta_N \Big(\tilde{\phi}(p)\Big)$ is exactly what one obtains by applying the coproduct rule - cf. (\ref{ow}) -  to the momentum-dependent expressions
\begin{equation}
\phi (p) = \tilde{\phi}(p) = \frac{i h}{8} \Big( e^{i \frac{p}{2}} + e^{-i \frac{p}{2}}\Big).
\end{equation}
\hfill{$\square$}

\medskip

We would like to end this section by noticing the similarities and the differences of the coproduct (\ref{deltaJ}) for the boost generator, with the coproduct of the form (\ref{secret}) typically associated to the {\it secret-symmetry} generator. The structure is of course very similar, but we can see that there are two main differences. One is the type of deformation supported by the tail, which of course directly follows from the coproduct of the level-zero supercharges. The biggest difference is, however, that the secret generator has a matrix representation on the magnon excitations, where it acts as the fermionic number times a momentum-dependent scalar factor. This is not the case for the boost generator, which does not commute with the central elements, and acts therefore as a differential operator in the momentum (or the rapidity) variable. This will be further exploited in the last section of the paper.

\subsection{Classical limit}
We can obtain a classical limit of the massles $R$-matrix (\ref{eq:RLL}). In traditional settings, this would lead to the notion of the so-called {\it classical $r$-matrix}, which is the first-order term in the expansion of the quantum $R$-matrix around the identiy. In the theory of quasi-triangular Hopf algebras, the classical $r$-matrix holds a special value, as it can often uniquely characterise the all-order expansion ({\it Belavin-Drinfeld theorems}, cf. \cite{BelavinDrinfeld1}).  

Because of the negative sign in the all-fermion entry, we shall however first need to switch to a new matrix
\begin{equation}
\check{R} = \Pi_s \circ R,
\end{equation}
with $\Pi_s$ the graded permutation-operator on states of the tensor-product representation\footnote{Notice that this is different from taking $\Pi(R)$.}. We then expand it in the regime where
\begin{equation}
\label{smallp}
p_i \sim \hbar + k_i \hbar^2 + {\cal{O}}(\hbar^3)
\end{equation}
as
\begin{equation}
\label{expa}
\check{R} = \mathbbm{1} \otimes \mathbbm{1} + \, \hbar \, \check{r} + {\cal{O}}(\hbar^2).
\end{equation}
We obtain the following classical object\footnote{We are disregarding the contribution from the overall scalar factor $\Phi$ for this analysis.}:
\begin{eqnarray}
\label{eq:rLL}
&&\check{r} |\phi\rangle \otimes |\phi\rangle\ :=\ 0,\qquad \qquad \qquad \qquad \qquad \check{r} |\psi\rangle \otimes |\psi\rangle\ :=\ 0, \\ \nonumber
&&\check{r} |\phi\rangle \otimes |\psi\rangle\ :=\ \frac{1}{2} (k_2 - k_1) |\psi\rangle \otimes |\phi\rangle,\qquad
    \check{r} |\psi\rangle \otimes |\phi\rangle\ :=\ \frac{1}{2}(k_1-k_2) |\phi\rangle \otimes |\psi\rangle,
\end{eqnarray} 
which can simply be written as
\begin{equation}
\label{no}
\check{r} = \frac{1}{2} \, (k_2 - k_1)[E_{12} \otimes E_{21} + E_{21} \otimes E_{12}]
\end{equation}
in terms of the matrix-unities $E_{ij}$, which send state $|j\rangle$ into state $|i\rangle$.

Because of the need of taking the graded permutation on the outgoing states, (\ref{no}) is not skew-symmetric but rather {\it symmetric}\footnote{We recall the fermionic nature of $E_{12}$ and $E_{21}$.}, {\it i.e.} $\Pi(r)(k_2,k_1) = r_{12}(k_1,k_2)$. It also does not satisfy the standard classical Yang-Baxter equation, but rather a version of it with suitable permutations inserted. It therefore lies outside the Belavin-Drinfeld theory, but it might nevertheless be useful in correlation with the classical limit of the $q$-Poincar\'e coalgebra and its Poisson structure \cite{Ballesteros:1999ew}.

The classical limit of the algebra generators is finite if we simultaneously send the coupling constant $h$ to infinity, in such a way that
\begin{equation}
h \sim \hbar^{-1}, \qquad h \, p_i \sim 2 g^2,
\end{equation}
with $g$ a constant. One obtains (on the branch of the dispersion-relation which we have here chosen)
\begin{equation}\label{leftrepa}
\begin{gathered}
\alg{S}_{\smallR} \ \sim \ - g \begin{pmatrix}0&0\\1&0\end{pmatrix}~,\qquad
\alg{Q}_{\smallR} \ \sim \ - g \begin{pmatrix}0&1\\0&0\end{pmatrix}, \\
\alg{Q}_{\smallL} \ \sim \ g \begin{pmatrix}0&0\\1&0\end{pmatrix}~,\qquad
\alg{S}_{\smallL} \ \sim \ g \begin{pmatrix}0&1\\0&0\end{pmatrix}. 
\end{gathered}
\end{equation}
\begin{equation}\label{eq:defofexa}
H_{\smallL} = H_{\smallR} = - P = - K \sim g^2.
\end{equation}
It is interesting to notice that, in order to obtain the expression (\ref{expa}), we have to take the leading term of the small-$\hbar$ expansion (\ref{smallp}) to be a {\it constant} times $\hbar$, which differs from the standard near-BMN expansion \cite{Berenstein:2002jq}.

\section{Physical considerations}\label{phys}

Let us speculate on the possible physical meaning of the various structures which we have uncovered.

The first thing to notice, as already pointed out by \cite{Charles}, is that the generator $\alg{H}$,
whose eigenvalue on a one-particle state coincides with the single-excitation energy, is no  longer additive on two-particle states in this new framework, as $\Delta_N$ acts non-trivially on both $\alg{H}_{\smallL}$ and $\alg{H}_{\small_R}$. Nevertheless, by direct calculation using (\ref{leftrep}), one finds
\begin{eqnarray}
\label{umklapp}
\Delta_N(\alg{H}_{\smallL}) = \Delta_N(\alg{H}_{\smallR})\to h \, \sin \frac{p_1 + p_2}{2} \, \mathbbm{1} \otimes \mathbbm{1}, \qquad \Delta_N(\alg{H}) \to 2 h \, \sin \frac{p_1 + p_2}{2}\, \mathbbm{1} \otimes \mathbbm{1}, 
\end{eqnarray}
where the arrow denotes the fact that we are taking a specific representation of the left-hand side.
On the one hand, this explicitly shows that the coproduct $\Delta_N$ is co-commutative on all central charges, which is indeed a necessary condition for the existence of an $R$-matrix. On the other hand, it echoes two physical phenomena which we will recall here below, and which will shed a new light on the problem of massless string-mode scattering.

\subsection{Phonon analogue} 

In phonon physics, (\ref{umklapp}) can be seen to describe the so-called {\it umklapp} scattering. This was elucidated in \cite{Ballesteros:1999ew,Ballesteros2,Bonechi:1992qf}. Let us remark that the dispersion-relation (\ref{eq:defofex}) coincides on the chosen branch\footnote{Taking into account the modulus which characterises its general expression, one observes that the $AdS_3$ massless dispersion-relation {\it exactly} coincides with the one for phonons.} with the single-particle dispersion-relation of a phonon on a one-dimensional harmonic chain of lattice spacing $h^{-1}$ (and rescaled momentum): 
\begin{equation}
E(p) = 2 h \, \Big|\sin \frac{p}{2}\Big|.
\end{equation}
In an {\it umklapp} process, two incoming phonons of momenta $p_1$ and $p_2$ merge into a third outgoing phonon, with the energy-conservation condition given by
\begin{equation}
\label{bound}
E_1 + E_2 = 2 h \, \Big|\sin \frac{p_1}{2}\Big| + 2 h \, \Big|\sin \frac{p_2}{2}\Big| = 2 h \, \Big|\sin \frac{p_1 + p_2}{2}\Big| = E_3.
\end{equation} 
Of course, this process selects very specific momenta. Momentum conservation supplements the energy condition, but it leaves the ambiguity of adding to the momentum $p_3$ of the outgoing particle an integer multiple of the reciprocal lattice vector
\begin{equation}
\label{umk}
p_1 + p_2 = p_3 + 2 n \pi, \qquad n \in \mathbbmss{Z}
\end{equation}  
which respects (\ref{bound}). In this way, the outgoing momentum, as well as the incoming ones, can always be chosen to belong to the Brillouin zone: $p_i \in [- \pi, \pi), \, i=1,2,3$. The term {\it umklapp} (from the German {\it flip-over}) specifically refers to the case $n \neq 0$ in (\ref{umk}), when the outgoing momentum $p_3$ might for instance be much smaller than the sum of the two incoming ones.

In our setting, this process is forbidden by integrability. Nevertheless, 
\begin{equation}
\Delta_N(\alg{H}) = 2 h \, \sin \frac{p_1 + p_2}{2}\, \mathbbm{1} \otimes \mathbbm{1}
\end{equation}
is conserved, simply because of momentum conservation. The point is that the new coproduct allows instead a novel interpretation of the scattering problem, precisely because of the special way it acts on all the central charges. In fact, one has 
\begin{eqnarray}
\label{compo}
&&\Delta_N(p) = (p_1 + p_2) \, \mathbbm{1}_\otimes = p_3 \, \mathbbm{1}_\otimes, \qquad \Delta_N(\alg{H}) = 2 h \sin \frac{p_3}{2} \, \mathbbm{1}_\otimes\ \qquad \mathbbm{1}_\otimes \equiv \mathbbm{1} \otimes \mathbbm{1}.
\end{eqnarray}
The formulas (\ref{compo}) exactly coincide with the quantum numbers of a {\it single} massless particle of momentum $p_3 = p_1 + p_2$ and energy $E(p_3)$. It is only thanks to these special relations that we can simply think of the scattering process as a free propagation of a massless compound, having the right energy {\it vs} total-momentum dispersion. This is in line with the observation of \cite{Bonechi:1992qf}. The $q$-deformed Poincar\'e symmetry implies that the scattering particles move like a single phonon. In our case, one might find useful to define a global (momentum dependent) {\it velocity parameter} $v_T$, given by
\begin{eqnarray}
E = v_T  P
\end{eqnarray}
(which tends to $e_0 = c p_1$ in the undeformed relativistic limit).

\subsection{Singularities}

In the context of integrable systems, the coproduct $\Delta_N(\alg{H})$ is rather interpreted as a potential bound-state condition \cite{Bonechi:1992cb}. In the case of magnons on a ferromagnetic Heisenberg spin-chain, which have a different dispersion-relation from the one we are dealing with, this was explicitly shown in \cite{Bonechi:1992cb}. In that context, the condition precisely matches the presence of a simple pole in the magnon $S$-matrix, corresponding to a bound-state solution of the Bethe ansatz equations. 

In our context, bound-state formation is prohibited \cite{upcom}, in accordance with general expectations of massless scattering \cite{Zamo}. Nevertheless, this coproduct might turn out to be useful to detect singularities of the $S$-matrix in the complex-momentum plane.

There is also another context where so-called {\it umklapp terms} appear, and that is in relation with interactions in effective Hamiltonians which describe spin-chains in definite regimes of the parameters (see for example \cite{Sirker}). Our findings might eventually be useful, were one to try and repeat a similar type of effective analysis for the $AdS_3$ spin-chain / sigma model.

\section{Representation for the boost operator}

A final comment we would like to make is that the boost operator of the $q$-deformed algebra can be used to uniformise the dispersion-relation \cite{Gomez:2007zr}. In fact, we can fulfil the commutation relations (\ref{algeq}) in our representation by resorting to differential operators in one complex variable $z$, whereby 
\begin{equation}
\alg{J}_{\smallL} = \alg{J}_{\smallR} = \frac{1}{\sqrt{\mu}} \, \partial_z.
\end{equation} 
In this fashion, we simply need to solve
\begin{equation}
\frac{1}{\sqrt{\mu}}  \, \partial_z \, p(z) = \, i h  \sin \frac{p(z)}{2} \qquad \mbox{hence} \qquad \partial_z \, p(z) = \, 2 i  \sin \frac{p(z)}{2}.
\end{equation}
This has the following solution (setting the integration constant to a convenient value):
\begin{equation}
p(z) = 4 \cot^{-1} \exp (- i z).
\end{equation}
Such a solution is also consistent with the double commutator we deduce from  (\ref{algeq}), {\it i.e.}
\begin{equation}
[\alg{J}_A,[\alg{J}_B,p]] = i \frac{e^{i p} - e^{-i p}}{2 \mu} \qquad \mbox{hence} \qquad \partial_z^2 \, p(z) = - \sin p(z),
\end{equation} 
where {\footnotesize $(A,B) = (L,L), (L,R), (R,L), (R,R)$}. 
At this point, we can choose to work with this parameterisation. It is then easy to see that 
\begin{equation}
z \to z \pm \pi \, \qquad \mbox{implies} \qquad p \to - p.
\end{equation}
With a specific choice of analytic path, one of the above transformations could be used to implement crossing symmetry. 

Let us notice that, when $p$ becomes small, $z$ goes to infinity as a logarithm:

\begin{equation}
z = i \log \cot \frac{p}{4} \to - i \log p + i \log 4 - \frac{i p^2}{48} \qquad\mbox{as} \, \, \,\, \,  p \to 0.
\end{equation}

This means that, in the undeformed (relativistic ) limit (\ref{unde}), we have
\begin{equation}
z \to z_{lim} = - i \log \epsilon p_1, \qquad \mbox{hence} \qquad p_1 =  e^{\theta}, \qquad \theta = \lim_{\epsilon \to 0} \, (i z_{lim} - \log \epsilon)    
\end{equation}
which identifies $\theta$ with Zamolodchikov's massless relativistic rapidity \cite{Zamo}. 

Let us finish by noticing how, as remarked in \cite{Gomez:2007zr}, the boost generator has been used in the context of lattice models, to implement a discrete version of relativistic invariance. The spirit was again the one of retaining Poincar\'e symmetry as much as possible, and it brought to the concept of Baxter's {\it corner transfer matrix} (see \cite{Thacker} for an account). The $AdS/CFT$ dispersion-relation itself interestingly combines periodic lattice features with a relativistic flavour, something which becomes even more striking in the massless case. It might therefore be beneficial to explore this connection further.

\section{Comments and Future Directions}\label{comments}

In this paper we have found an exact realisation of a $q$-deformed Poincar\'e superalgebras in the massless integrable sector of the $AdS_3$ superstring. Such a deformed symmetry was partially manifest in $AdS_5$, and it was used to allow for alternative interpretations of the peculiar magnon dispersion-relation, and of the representation of the short magnon multiplet. For a suitable choice of branches, the massless dispersion-relation allows to overcome the obstacles of the massive case, and displays a host of remarkably natural features. Above all, the comultiplication on all central charges is simply obtained by acting trivially on the momentum generator. This was used in other contexts as a way of describing the so-called {\it umklapp} scattering of phonons, which our system bares a mysterious similarity to. 

We have studied the {\it boost} operator and its coproduct, demonstrating that it is an algebra homomorphism. The existence of this generator is probably the deepest novelty of this deformed setting. It facilitates the uniformisation of the dispersion by acting as a differential operator on the representation labels, realising a series of intriguing pseudo-relativistic phenomena. This is in the same spirit as in the theory of integrable lattice models, where the same object serves the purposes of a remnant of Poincar\'e invariance ({\it corner transfer matrix}). In this respect, it would be very interesting to investigate whether this generator has any worldsheet realisation, possibly in terms of an action on the classical fields of the string sigma model\footnote{We thank B. Stefa\'nski for pointing this out.}. Given the persistence of mismatches between the exact predictions and the perturbative results for the $AdS_3$ (and $AdS_2$) massless sector (cf. \cite{Sundin:2016gqe}), our analysis might provide an alternative way to approach the issue, by giving a new set of algebraic relations one could try and check using string perturbation-theory.  
 
One open question related to these findings is the following. It is not yet clear how to prove that the coproduct of the boost generator is a symmetry of the $S$-matrix, something which is instead easy to show for all the rest of the algebra. In its differential form, it seems plausible that it would act on the overall scalar factor as well, perhaps {\it fixing} it purely from symmetry. This is an exciting concept, but it also makes it incredibly hard to show. We also do not know how to represent such a differential action on {\it states}. 

In this respect, it might be useful to study the universal $R$-matrix of our algebra. Recently, a universal $R$-matrix for the Lie super-algebra of $AdS_5$, with a $q$-deformation super-imposed, has been found \cite{Beisert:2016qei}, and a crucial role of certain automorphisms is there firmly established. It is fascinating to consider the connection with the formalism of this paper, and the possibility of finding perhaps a common formulation of the two problems. In both cases actually, the extension to {\it quantum-affine} and {\it Yangian}, respectively, appears to be the first necessary step.

We plan to study these, and other related questions, in future work on this topic.

\section*{Acknowledgments}

We would like to warmly thank R. Borsato, O. Ohlsson Sax, A. Sfondrini and B. Stefa\'nski for illuminating discussions, and for reading the manuscript and providing extremely useful comments. We also much thank M. Abbott,  G. Festuccia, B. Hoare, M. de Leeuw, F. Nieri, A. Pittelli, A. Prinsloo, V. Regelskis and S. van Tongeren, for very important discussions. A.T. thanks
the EPSRC for funding under the First Grant project EP/K014412/1 {\em Exotic quantum groups,
Lie superalgebras and integrable systems}. The authors thank the STFC for support under the Consolidated
Grant project nr. ST/L000490/1 {\em Fundamental Implications of Fields, Strings and Gravity}.
A.T. also acknowledges useful conversations with the participants of the ESF and STFC supported
workshop {\em Permutations and Gauge String duality} (STFC - 4070083442, Queen Mary
U. of London, July 2014), and the organisers and participants of the meeting on {\it $\eta$ and $\lambda$ deformations in integrable systems and supergravity} workshop, Albert
Einstein Institute, Bern, and the ETH conference {\em All about $AdS_3$} (Switzerland, 2015), of the {\it Selected Topics in Theoretical High Energy Physics} conference in Tbilisi (Georgia, 2015) and of the Nordita program {\em Holography and Dualities 2016: New Advances in String and Gauge Theory}, for fruitful conversations and a stimulating atmosphere.

No data beyond those presented and cited in this work are needed to validate this study.

\end{document}